\documentstyle[aps,preprint]{revtex}

\input epsf

\begin{document}

\title{Fixed-Velocity Chiral Sum Rules for Nuclear Matter}

\author{Thomas D. Cohen}
\address{Department of Physics, University of Maryland, 
 College Park, MD 20742-4111, USA}

\author{Wojciech Broniowski}
\address{H. Niewodnicza\'nski Institute of Nuclear Physics,
            PL-31342 Krak\'ow, Poland}

\date{U.MD PP 97-077,~~DOE/ER/40762-109,~~INP 1753/PH ~~(February 1997)}

\maketitle

\begin{abstract}
Infinite sets of sum rules involving the excitations of infinite 
nuclear matter are derived using only completeness, the current algebra
implicit in QCD, and relativistic covariance. The sum rules can be
used for isospin-asymmetric nuclear matter, including neutron matter.
They relate the chiral condensate and the isospin density to weighted
sums over states with fixed velocity relative to the nuclear matter 
ground state.  
\end{abstract}


\bigskip

Sum rules have been an important tool in theoretical physics for
a very long time. Their importance stems in large measure from the
fact that they allow one to make concrete predictions concerning
sums of matrix elements even when one does not have, or cannot solve the
theory for the dynamics of the system. One interesting and important
system for which the underlying dynamics is known but
cannot presently be solved is the nuclear medium.  While we are
quite onfident that the underlying dynamics for the strong interactions is
QCD, it is quite unlikely that it will prove to be a tractable task to
calculate nuclear properties from QCD any time in the forseeable 
future.  This raises an obvious question, namely whether one can
use known properties of QCD, such as current algebra, to derive sum
rules for nuclear properties. 

In this work we will not discuss sum rules for
finite nuclei.  Instead, we concentrate on infinite nuclear matter with
arbitrary baryon density and arbitrary isospin density.  We do this
for several reasons.  One is  practical:  if the
system is translationally invariant (and hence necessarily infinite)
momentum is a good quantum number and can be used to label 
excitations.   A second reason is that interiors of
large nuclei are well approximated by nuclear matter and learning about
nuclear matter gives a more general insight into nuclear
systems than the study of individual nuclei.  A third reason is 
that the cores of neutron stars are essentially isospin-asymmetric
nuclear matter (neutron matter).  Finally, infinite 
nuclear matter is theoretically interesting in its own right.  There is 
obviously one major downside
to studying infinite nuclear matter: the connection between sum rules
for infinite nuclear systems  and experimental observables for
real nuclei is not immediately clear.

In previous works \cite{CB1,CB2} we have
studied the chiral properties of nuclear matter. 
That is, we examined the behavior of modes as the quark mass is formally
taken to zero.  We showed that certain zero-momentum modes had to go
to zero energy as the (current) quark masses were taken to zero.   
In the process of deriving this result we used current algebra 
and completeness to derive sum rules for the squares of the matrix 
elements of the divergence  of the axial current between zero momentum 
states. Analogous sum rules are also implicit in the work of 
Lutz, Steiner and Weise \cite{Lutz}. 

Now, in Refs. \cite{CB1,CB2} the central stress was on the fact that
these sum rules were necessarily saturated by modes whose energy went
to zero as the quark masses did.  However, it is worth observing that
these sum rules in no way depend on the chiral
limit.  As sum rules, they hold for {\it any value of the quark mass}. 
This is significant for two reasons.  The first is that compared
to typical nuclear mass scales, the effects of finite up and down quark
masses should not be regarded as small.  Moreover, there
has been considerable interest in kaonic excitations in dense nuclear
matter leading, for example, to kaon 
condensation \cite{KN}-\cite{waas}.
 
Accordingly, it is useful to study strange excitations, and the strange
quark mass is certainly not small on nuclear physics scales.
The fact that the sum rules hold for any value of the quark mass was 
used explicitly in Ref.~\cite{BH} in which the Nambu--Jona-Lasinio
model was studied at mean-field level and in which the contribution to
the sum rules coming from the zero-sound mode as well as from
pion modes ({\it i.e.} modes that map smoothly onto the vacuum pion
modes as the density goes to zero) were explicitly calculated.

There is one major drawback to the sum rules derived in
\cite{CB1,CB2}: they only give information about modes
with zero momentum.  This is unfortunate because one of the
most interesting issues about nuclear matter is the dependence of the
energy of a mode on its momentum. The previous sum rules
are completely insensitive to such effects. In the present paper,
we will show how to exploit relativistic covariance to derive sum rules
which are sensitive to states with nonzero momentum.  More precisely,
they involve sums over states with a fixed velocity,
$\beta = \vec{q}/E$,  where $\vec{q}$ and $E$ are momentum 
and energy of the state, 
measured relative to the ground state of nuclear matter.

We begin by stating the result in
the rest frame of the medium; a covariant formulation will be given
later. Also, for simplicity the notation is for two flavors. 
The generalization to three
flavors is straightforward. Let us denote the ground state of 
{\em translationally invariant} nuclear medium by $|C \rangle$.
The state is subject to space-independent 
constraints which fix its baryon density, $\rho_B$, and 
isospin density, $\rho_{I=1}$.
We are concerned with the spectrum of excitations with quantum
numbers of the pion (analogous results apply for the  kaon) on top of the 
nuclear medium. These excited states are denoted by $|j_a, \vec{q} \rangle$, 
where $j_a$ labels modes with isospin equal ($a=0$) , greater by one
unit ($a=+$), or lower by one unit ($a=-$) than the 
isospin of the state $|C \rangle$, and $\vec{q}$ is the momentum
of the mode.  
The excitation energy of the mode, {\em i.e.}, 
the difference of its energy and the reference energy of the state 
$|C \rangle$, is denoted by $E_{j_a}(\vec{q})$. We stress that no
assumptions are made as to what the excited states 
$|j_a, \vec{q} \rangle$ are; they include collective 
excitations, one-particle--one-hole continuum,
two-particle--two-hole continuum, {\em etc.}---in short, all states 
that have quantum numbers of the pion.

We define the following spectral densities associated with the 
divergence of the axial current $A^a_\mu$: 
\begin{equation}
\label{spect}
\sigma^a_C(E,\vec{q}) = \sum_{j_a} \frac{\left | 
\langle j_a, \vec{q} | \partial \cdot A^a(0) | C \rangle \right |^2}
{2|E_{j_a}(\vec{q})|} \delta(E-E_{j_a}(\vec{q})) \;, \;\;\;\;\; a=0,+,- \;. 
\end{equation}
In general, the sum is over continuum states and assumes
the form of an integral. 
We will show that three classes of
sum rules exist, which relate quark condensates and the isovector
density to spectral integrals over $\sigma^a_C$: 
\newcounter{aux}
\setcounter{aux}{\arabic{equation}}
\renewcommand{\theequation}{\roman{equation}}
\setcounter{equation}{0}
\begin{eqnarray}
\label{sum:I}
&& \!\!\!\!\!\!\!\!\!\!\!\!\!
- m_u \langle \overline{u}u \rangle_C 
- m_d \langle \overline{d}d \rangle_C 
= 2 \int_{0}^{\infty} 
 \frac{dE}{E}{\sigma^0_C}(E, \vec{q}=\vec{\beta}E) \;, \\
\label{sum:II}
&& \!\!\!\!\!\!\!\!\!\!\!\!\!
 - (m_u+m_d) \langle \overline{u}u + \overline{d}d \rangle_C 
= 2 \int_{- \infty}^{\infty} \frac{dE}{E} \left (
{\sigma^+_C}(E, \vec{q}=\vec{\beta}E) + 
{\sigma^-_C}(E, \vec{q}=\vec{\beta}E) 
\right )\;, \\
\label{sum:III}
&& \!\!\!\!\!\!\!\!\!\!\!\!\!
- \rho_{I=1} =  
\int_{- \infty}^{\infty} \frac{dE}{E^2} \left (
{\sigma^+_C}(E, \vec{q}=\vec{\beta}E) - 
{\sigma^-_C}(E, \vec{q}=\vec{\beta}E) 
\right ) \;, 
\end{eqnarray}
\setcounter{equation}{\arabic{aux}}
\renewcommand{\theequation}{\arabic{equation}} 
Note that the spectral functions $\sigma^a_C$ in the integrands
have the arguments constrained to $\vec{q}/E = \vec{\beta}$. This means
that the modes contributing to the sum rule move (in the rest
frame of the medium) with the fixed velocity $\vec{\beta}$. 

\begin{figure}
\epsfxsize = 12 cm
\centerline{\epsfbox{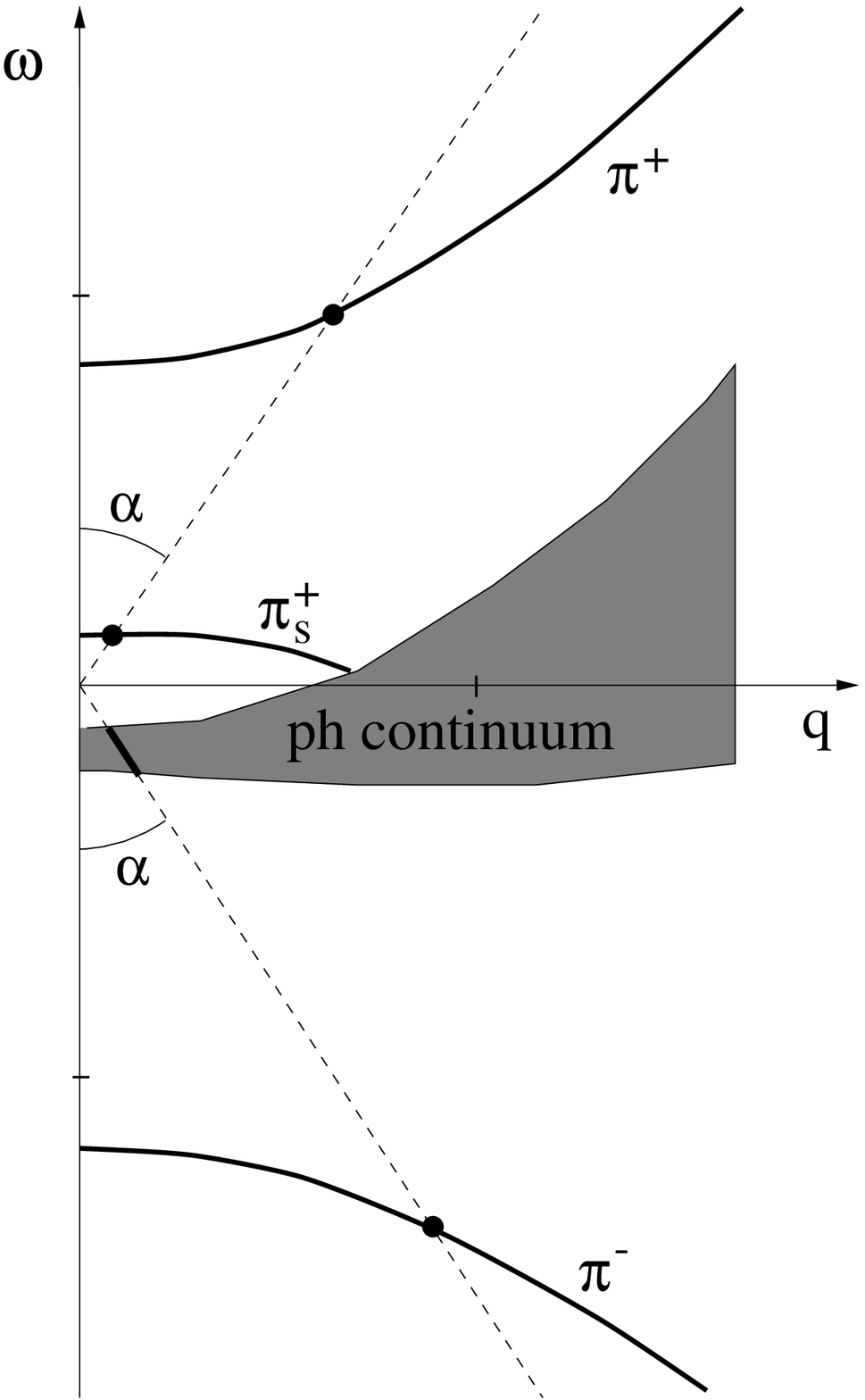}}
\caption{ }
\label{fig:schem}
\end{figure}

Figure \ref{fig:schem} illustrates schematically the described situation.
What is shown is a typical result of a nuclear calculation
of the charged pionic excitation spectrum in neutron matter up to the
one-particle--one-hole level
\cite{rhowil}. Solid lines correspond to the
$\pi^+$ and $\pi^-$ poles, and the shaded region 
represents the one-particle--one-hole continuum. In addition,  a
possible spin-isospin sound mode $\pi^+_s$ \cite{rhowil} 
is plotted. On the vertical axis is the frequency
of the mode, $\omega$. The energy of the excitation 
which positive (negative) isospin modes
the energy is equal to $+$ ($-$) $\omega$. The dashed line shows
the integration path in the sum rules. It corresponds to
$|\vec{q}|/E = \beta$, {\em i.e.}, to modes moving with the fixed
velocity $\beta$ with respect to the medium (which is at rest).
Contributions to the sum rules coming from various
excitations are denoted by blobs (poles) and the thick line (the 
particle-hole cut).
The dashed line is inclined to the vertical axis at the angle 
$\alpha = {\rm arctg} \beta$. At various values of the velocity
$\beta$ different regions of the spectral density are sampled. Note
that $\alpha \le 45^o$. One should note that in reality
the corresponding figure would be much more complicated. 
Due to multi-particle--multi-hole
continua spanning the whole range of $\omega$ and $\vec{q}$, 
all states have finite widths, and contributions to sum
rules are collected from everywhere along the dashed lines.

We now pass to the proof of sum rules (\ref{sum:I})-(\ref{sum:III}). We
use two facts. Firstly, the chiral current algebra 
satisfied by QCD yields the following operator identities:
\begin{eqnarray}
\label{id1}
[Q_5^a, [Q_5^a, {\cal H}(0)]] &=&
\overline{q}(0)\{ \tau^a/2,
\{\tau^a/2,M\}\} q(0) \;, \;\;\;\;\; {\rm any~} a=1,2,3 \;, 
\\ \label{id2}
[Q_5^a, A_0^b(0)] &=& i \epsilon^{abc} V_0^c(0) \;, 
\end{eqnarray}  
where $V^a_\mu(x) = \overline{q}(x) \gamma_\mu \frac{1}{2} \tau^a q(x)$ 
and $A^a_\mu(x) = \overline{q}(x) \gamma_5 \gamma_\mu \frac{1}{2} \tau^a q(x)$ 
are vector and axial currents,
$Q_5^a = \int d^3x\,A^a_0(x)$, 
${\cal H}(x)$ is the QCD Hamiltonian density, and
$M = {\rm diag}(m_u, m_d)$ is the quark mass matrix.

Secondly, we assume that the medium is 
{\em translationally invariant}. This is true for infinite-volume 
nuclear matter. The important observation is, however, that the medium
need not be at rest. Medium moving with a constant velocity
$-\vec{\beta}$ in a reference frame 
is also translationally invariant. We denote such a state 
by $| C, -\vec{\beta} \rangle$.
The sum rules are constructed in the usual way \cite{adler}. The  
identities (\ref{id1})-(\ref{id2}) are sandwiched by 
the state $|C, -\vec{\beta} \rangle$. Inside the LHS we insert 
covariantly normalized
intermediate states \cite{CB1}, using the identity
\begin{equation}
\label{inter}
1 = \sum_j \int \frac{d^3p}{(2 \pi)^3 2 |E_j^{(\beta)}(\vec{p})|}
 |j, \vec{p} \rangle \langle j, \vec{p}| \;.
\end{equation}
The intermediate states can be labeled by momentum $\vec{p}$ since 
$|C, -\vec{\beta} \rangle$ is translationally invariant. Index $j$
sums over all additional quantum numbers. The excitation
energy $E_j^{(\beta)}(\vec{p})$ is the difference of the energy of the 
excited state and the state of the moving medium, {\em i.e.}, we have 
\begin{equation}
\label{energy} 
H |C, -\vec{\beta} \rangle = 
 E_C^{(\beta)} |C, -\vec{\beta} \rangle \;, \;\;\;\;\; 
H |j, \vec{p} \rangle = 
\left ( E_C^{(\beta)} + E_j^{(\beta)}(\vec{p}) 
\right ) |j, \vec{p} \rangle \;.
\end{equation}
The momentum $p$ is defined analogously---it is the momentum relative
to the ground state of nuclear matter.
Note that $E_j^{(\beta)}(\vec{p})$ and $\vec{p}$ form
a Lorentz four-vector.
The immediate result of the described 
construction is the following set of sum rules:
\begin{eqnarray}
\label{sumbeta:I}
&&\!\!\!\!\!\!\!\!\!\!\!\!\!
- \langle C, -\vec{\beta}| ( m_u \overline{u}u + m_d  \overline{d}d )
 |C,-\vec{\beta} \rangle 
= \sum_{j_0} 
\frac{| \langle j_0, \vec{p}=0 | \partial \cdot A^0(0) 
| C, -\vec{\beta} \rangle |^2}
{| E^{(\beta)}_{j_0}(\vec{p}=0) | E^{(\beta)}_{j_0}(\vec{p}=0)}\;, \\
\label{sumbeta:II}
\vspace*{.15in}
&&\!\!\!\!\!\!\!\!\!\!\!\!\!
 - \langle C, -\vec{\beta}|(m_u+m_d)(\overline{u}u + \overline{d}d)
 |C,-\vec{\beta} \rangle 
= \sum_{j_+} 
\frac{| \langle j_+, \vec{p}=0 | \partial \cdot A^+(0) 
| C, -\vec{\beta} \rangle |^2}
{| E^{(\beta)}_{j_+}(\vec{p}=0) | E^{(\beta)}_{j_+}(\vec{p}=0)} \nonumber \\
\vspace*{.12in}
&&\;\;\;\;\;\;\;\;\;\;\;\;\;\;\;\;\;\;\;
\;\;\;\;\;\;\;\;\;\;\;\;\;\;\;\;\;\;\;\;\;\;\;\; + \sum_{j_-} 
\frac{| \langle j_-, \vec{p}=0 | \partial \cdot A^-(0) 
| C, -\vec{\beta} \rangle |^2}
{| E^{(\beta)}_{j_-}(\vec{p}=0) | E^{(\beta)}_{j_-}(\vec{p}=0)} \;, \\
\label{sumbeta:III}
\vspace*{.15in}
&&\!\!\!\!\!\!\!\!\!\!\!\!\!
- (1  -   \beta^2)^{-1/2} \, \rho_{I=1}  
= \sum_{j_+} 
\frac{| \langle j_+, \vec{p}=0 | \partial \cdot A^+_0(0) 
| C, -\vec{\beta} \rangle |^2}
{| E^{(\beta)}_{j_+}(\vec{p}=0) |^3} \nonumber \\ 
\vspace*{.12in}
&&\;\;\;\;\;\;\;\;\;\;\;\;\;\;\;\;\;\;\;
\;\;\;\;\;\;\;\;\;\;\;\;\;\;\;\;\;\;\;\;\;\;\;\; - \sum_{j_-} 
\frac{| \langle j_-, \vec{p}=0 | \partial \cdot A^-_0(0) 
| C, -\vec{\beta} \rangle |^2}
{| E^{(\beta)}_{j_-}(\vec{p}=0) |^3} \;.
\end{eqnarray}
Sum rules (\ref{sumbeta:I})-(\ref{sumbeta:II}) are the consequence of 
Eq.~(\ref{id1}), and sum rule (\ref{sumbeta:III}) follows from
Eq.~(\ref{id2}). The factor $(1 - \beta^2)^{-1/2}$ is the dilatation
factor for the isospin density, which is the time-component of a Lorentz
four-vector (our notation is that $\rho_{I=1}$ is the isospin density
in the rest frame on the medium). 
Note that only states with $\vec{p}=0$ contribute to the above sum rules.

We can now make a boost with velocity $\vec{\beta}$ to the rest frame of the
nuclear matter. This boost transforms
Eqs.~(\ref{sumbeta:I})-(\ref{sumbeta:III}) into 
Eqs.~(\ref{sum:I})-(\ref{sum:III}). After this boost, 
the medium is at rest, and the excitations move with the fixed
velocity $\vec{\beta}$. 
We can write the sum rules in the covariant form.
Note there are two Lorentz vectors available for constructing
invariants: the four-momentum $p^\mu$ 
formed by the excitation energy and the 
momentum of the excited mode relative to the medium, and the
four-velocity of the medium, $u^\mu$. Introducing  
$s=p_\mu p^\mu$ and $y=p_\mu u^\mu$ we can rewrite
(\ref{sum:I})-(\ref{sum:III}) as
\newcounter{aux2}
\setcounter{aux2}{\arabic{equation}}
\renewcommand{\theequation}{\Roman{equation}}
\setcounter{equation}{0}
\begin{eqnarray}
\label{sumy:I}
&& \!\!\!\!\!\!\!\!\!\!\!\!\!\!\!
- m_u \langle \overline{u}u \rangle_C 
- m_d \langle \overline{d}d \rangle_C 
= 2 \!\! \int_{0}^{\infty} 
 \frac{dy}{y}{\sigma^0_C}(s=y^2  (1- \beta^2)^2 , y) \;, \\
\label{sumy:II}
&& \!\!\!\!\!\!\!\!\!\!\!\!\!\!\!
 - (m_u+m_d) \langle \overline{u}u + \overline{d}d \rangle_C 
= 2 \!\! \int_{- \infty}^{\infty} \frac{dy}{y} \left (
{\sigma^+_C}(s=y^2  (1- \beta^2) , y) + 
{\sigma^-_C}(s=y^2  (1- \beta^2) , y) 
\right ), \\
\label{sumy:III}
&& \!\!\!\!\!\!\!\!\!\!\!\!\!\!\!
- \langle V_\mu^0(0) \rangle_C = u_\mu \!\! 
\int_{- \infty}^{\infty} \frac{dy}{y^2} \left (
{\sigma^+_C}(s=y^2 (1- \beta^2), y) - 
{\sigma^-_C}(s=y^2  (1- \beta^2), y) 
\right ) \;, 
\end{eqnarray}
\setcounter{equation}{\arabic{aux2}}
\renewcommand{\theequation}{\arabic{equation}} 
\noindent Covariant forms of the spectral densities (\ref{spect}) appear in
the above equations. However, the arguments are constrained to
$s=y^2 \,  (1- \beta^2)$  where $\vec{\beta}$ is the velocity of the modes
in the medium's rest frame.   An alternative---and
equivalent---covariant formulation would be to  use $s$ as the 
integration variable in (\ref{sumy:I})-(\ref{sumy:III}), with $y$
constrained.  Such a formulation is in some sense more natural in that
it has an obvious vacuum limit---in the vacuum the spectral density is
independent of $y$ and the sum rules are written in terms of 
integrals over $s$. However, in the medium this is slightly awkward 
to do since negative energy states relative to the nuclear matter are
possible---{\it e.g.}, in neutron matter a state with $\pi^+$ quantum
numbers may lie below the neutron matter ground state---and hence $s$
can be a multi-valued function of $E$ and the integral must be extended
over each branch (see comment (4) below).

A few remarks are in place: 

(1) If $|C \rangle$ is the vacuum, 
then (\ref{sum:I})-(\ref{sum:II}) become the Gell-Mann--Oakes--Renner
sum rule \cite{GMOR}. In the chiral limit and for 
$m_u = m_d \equiv \overline{m}$  one obtains the familiar relation 
\mbox{$\overline{m} \langle \overline{q}q \rangle  
= F_{\pi}^2 m_{\pi}^2$}. Note, however, that the sum rules are exact
for any values of $m_u$ and $m_d$, also far away from the chiral limit.

(2) Sum rule (\ref{sum:III}) or (\ref{sumy:III}) is trivial for the
vacuum, and also for isospin-symmetric nuclear matter. However, it is
nontrivial for a medium which breaks the isospin symmetry. Its form
is reminiscent of  the sum rules of Fubini and Furlan (see {\em e.g.} 
\cite{adler}).

(3) In the vacuum the sum rules for 
various values of $\beta$ are equivalent. 
This is because the dispersion relations for pionic excitations are
fixed by
Lorentz invariance, {\it i.e.} \mbox{$E=\sqrt{q^2 + m_\pi^2}$}. 
This is no longer true in the presence of the medium,
and sum rules with different $\beta$ are physically distinct.   The
point is that if one were to apply one of our finite velocity sum rules
on a Lorentz-invariant state one would find that the size of the contributions
to the sum from any given mode  would not depend on $\beta$. In
contrast, since the medium breaks Lorentz invariance one finds that
the size of the contribution from a given mode to the sum does, in
general, depend on the value of $\beta$. In this sense these finite
velocity sum rules provide additional information about the spectrum
which is not present in the $\beta=0$ sum rules of 
Refs. \cite{CB1}-\cite{Lutz}.

(4) The integration variable in the sum rule for the $a=0$ excitations,
(\ref{sum:I}) or (\ref{sumy:I}), ranges from 0 to $\infty$, which
reflects the fact that the state $|C \rangle$ is the ground state 
of the matter subject to constraints. Therefore all excitations
within the constrained space have to raise the energy of the system.
Hence, the integration variables in (\ref{sum:I}) and (\ref{sumy:I})
are positive. This is not true for excitations with $a=\pm$, which
take the system out of the constrained space \cite{CB1}. In that case,
states $|j_{\pm}, \vec{q} \rangle$ may have lower 
energy than $|C \rangle$, and the integration in 
(\ref{sum:II})-(\ref{sum:III}) and (\ref{sumy:II})-(\ref{sumy:III}) has to
range from $-\infty$ to $\infty$. A model example of such a behavior is
given in \cite{BH}.   It is for this reason that it is awkward to
write the covariant versions of these sum rules over the $s$ variable.
  
(5) In models where the  pion fields satisfy the partially-conserved
axial current condition, {\em i.e.} where 
$\partial \cdot A^a = - m_\pi^2 F_\pi \pi^a$, the spectral 
densities (\ref{spect}) are proportional to the imaginary parts of the
pion propagator.  For such models  the sum rules can be written as
dispersion relations for the in-medium pion propagator.  

(6) The final remark concerns renormalization. Strictly speaking, 
sum rules (\ref{sum:I})-(\ref{sum:II}) or (\ref{sumy:I})-(\ref{sumy:II})
are ill defined on both sides.  The left-hand side has an ill-defined 
composite operator and the right-hand side has a divergent sum.
In order to make sense of the sum rules
one needs to define some scheme to renormalize the 
$\overline{q}q$ operator, as well as a subtraction term for the 
spectral sums.   At first blush this seems to suggest that these sum
rules are useless.  However, the
need for renormalization stems from the vacuum sector of the theory.
No new divergences are induced from the presence of the medium.   
Thus, once the renormalization is carried for the vacuum sector, it
holds also for the medium. This means, that the sum rules 
(\ref{sum:I})-(\ref{sum:II}) and (\ref{sumy:I})-(\ref{sumy:II}) should
be regarded as vacuum-subtracted sum rules, involving the difference
of the in-medium and the vacuum values of the quark condensates, 
\mbox{$\langle C | \overline{q}q | C \rangle - 
\langle {\rm vac} | \overline{q}q | {\rm vac} \rangle$}, and accordingly 
subtracted spectral densities.  
The third sum rule, (\ref{sum:III}) or (\ref{sumy:III}) involves
a conserved current, and as such requires no subtractions.

\begin{figure}
\epsfxsize = 12 cm
\centerline{\epsfbox{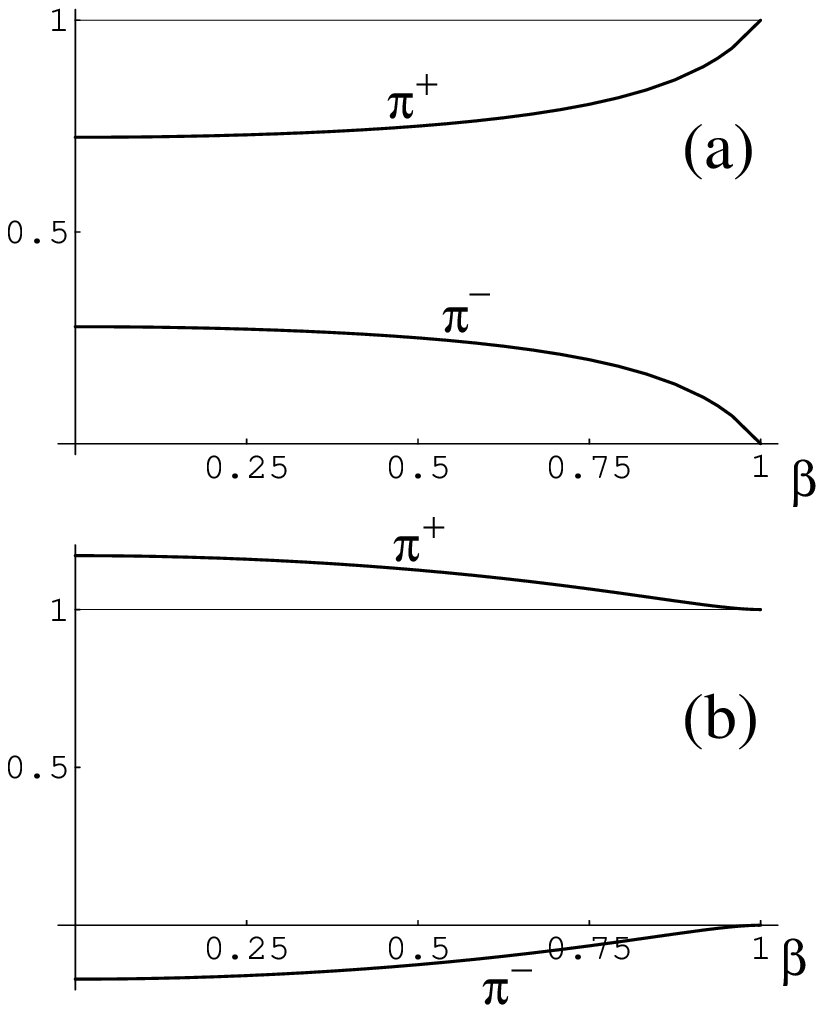}}
\caption{ }
\label{fig:sr}
\end{figure}

To make things concrete, we will illustrate how our sum rules work out
for  a simple toy model \cite{CB1}. 
The model describes the pion moving in an isospin-asymmetric medium
and interacting with it only via $\rho$-meson exchange.  Such a model
is obviously quite unrealistic since, among other things, it assumes that
the $\pi$-N coupling constant, and hence $g_A$ is strictly zero.  We set
$m_u=m_d=\overline{m}$.  In order to ensure chiral symmetry in such a model
one must  assume universal coupling of the rho meson and the KSFR
relation \cite{KSFR}, $ 2 g_\rho^2 \, = \, m_\rho^2/F_\pi^2$.
The inverse-charged pion propagator in the rest frame of the
medium has the form
\mbox{\mbox{$ G^\pm(q_0, \vec{q}) = 
q_0^2 \mp \rho_{I=1}/F_\pi^2 q_0 - \vec{q}^2 - m_\pi^2$}.}
The model has the following spectrum of neutral, 
positive and negative isospin excitations:
\begin{eqnarray}
\label{toy:spect}
E^0(\vec{q}) &=& \sqrt{m_\pi^2 + \vec{q}^2} \;, \nonumber \\ 
E^+(\vec{q}) &=& + \rho_{I=1}/(2 F_\pi^2) + 
 \sqrt{(\rho_{I=1}/(2 F_\pi^2))^2 + m_\pi^2 + \vec{q}^2} \;, \nonumber \\ 
E^-(\vec{q}) &=& - \rho_{I=1}/(2 F_\pi^2) + 
 \sqrt{(\rho_{I=1}/(2 F_\pi^2))^2 + m_\pi^2 + \vec{q}^2} \;.
\end{eqnarray}
After some simple algebra, sum rules (\ref{sum:I})-(\ref{sum:III}) 
can be cast in the form 
\begin{eqnarray}
\label{toy:sums}
\label{toy:I}
&& -\overline{m} \langle \overline{q} q \rangle_C = F_\pi^2 m_\pi^2 \;, \\
\label{toy:II}
&& -\overline{m} \langle \overline{q} q \rangle_C = 
F_\pi^2 m_\pi^2 \left [ \frac{a_-(\beta)}{a_-(\beta) + a_+(\beta)} + 
\frac{a_+(\beta)}{a_-(\beta) + a_+(\beta)} \right ] 
\equiv F_\pi^2 m_\pi^2 \;, \\
\label{toy:III}
&& -\rho_{I=1} = 
-\rho_{I=1} \left [ \frac{a_-(\beta)^2}{a_-(\beta)^2 - a_+(\beta)^2} - 
\frac{a_+(\beta)^2}{a_-(\beta)^2 - a_+(\beta)^2} \right ] 
\equiv -\rho_{I=1}  \;,
\end{eqnarray}
where $a_\pm(\beta) = \pm \rho_{I=1}/(2 F_\pi^2) + 
 \sqrt{(\rho_{I=1}/(2 F_\pi^2))^2 + m_\pi^2 (1-\beta^2)}$. 
Sum rule (\ref{sum:I}) acquires the trivial form (\ref{toy:I}).
Positive (negative) isospin contributions to the sum 
rules (\ref{sum:II}) and (\ref{sum:III}) correspond to first (second)
terms in the brackets in Eqs.~(\ref{toy:II})-(\ref{toy:III}). The
relative contribution of the positive and negative isospin modes 
depends on the value of $\beta$. These relative
contributions to the sum rules are plotted in Fig.~\ref{fig:sr}. The
figure is drawn for negative $\rho_{I=1}$ (neutron matter).   Note
that as $\beta$ changes, the relative weight of the modes changes.
This illustrates our central point---for translationally-invariant
systems  which break Lorentz invariance, these finite velocity sum
rules are inequivalent to their zero-velocity counterparts.  
As $\beta \to 1$, the positive isospin mode saturates both sum
rules. Also note that the two modes saturate the sum rules 
(\ref{sum:II})-(\ref{sum:III}), since there are no other excitation in
this model. 

Obviously, the toy model is not realistic, but it
illustrates our general statements: distinct sum rules for different
velocities $\beta$ and the exactness of the sum rules for any value of
$\overline{m}$.  
A more realistic model would involve at least
one-particle--one-hole-excitations. Cuts associated
with such excitations contribute to sum rules in addition to 
poles. An example of such a model can be found in Ref.~\cite{BH},
where it was found that the contribution of cuts to the sum rules 
(at $\beta=0$) was 
very small. In reality, as already mentioned in the discussion of 
Fig.~\ref{fig:schem}, there are in addition multi-particle--multi-hole
continua; hence, all quasiparticles acquire finite
widths. 

While the discussion heretofore has focused on pions in isospin
asymmetric nuclear matter, it should be obvious that the
results go over {\it mutis mutandus} to the study of excitations with
kaonic quantum numbers.  The key point is that these sum rules 
work for any quark mass.  Thus one can simply replace up quarks 
by strange quarks everywhere in the sum rules (or alternatively down
quarks by strange quarks).  
The only change is that instead of the isospin density 
one will have the u-spin or v-spin density on the left-hand side of 
sum rule (iii) or (III).  This is of significance since there has been
considerable interest in kaons in nuclear matter and in the possibility of
kaon condensation.

It is worth noting that although these sum rules give information 
about the spectrum away from $p=0$, they nevertheless do not make
contact with all possible modes.  The reason is that these modes are
fixed velocity with $\beta = p/E \le 1$ .  Now one should notice the
$p/E$ is the velocity of the mode and not its group velocity.  There
is nothing in principle to prevent the existence of
modes with $\beta >1$.  Our sum rules however tell us nothing about
such modes.  Recall, for instance, the p-wave pion
condensation.  If this were to happen,  as one approached  the
transition density a mode with  finite $p$ would approach 
zero energy.  Such modes do not contribute to our sum rules.

Finally, we should discuss how our sum rules might be useful.  Since
they are not based on finite nuclei, it is hard to see 
what the sum rules can tell us about experiments directly.  On the
other hand, the sum rules do provide very strong constraints on model
building.  After all, these sum rules were derived from very basic 
properties known to be satisfied by QCD---notably current algebra. 
Thus, any model which satisfies the various chiral Ward
identities must satisfy our sum rules.   Of course, it is very simple to 
construct models which satisfy the Ward identities.   For example, 
mean-field models based on chiral langrangians will.   However,
such models typically exclude much essential nuclear phenomenology such
as the effects of short-range
correlations and the need to put in form factors to cut off spurious
high-momentum physics.  A formalism including such effects may well
violate the Ward identities---that is to say, the approximations used to
make the calculation tractable while including these effects  will in
general not be  symmetry-conserving. One obvious use of our sum rules
is to test how badly the symmetries are violated.

WB would like to thank the TQHN group 
at the University of Maryland for its hospitality.
Research supported by the NSF--Polish Academy of Sciences
grant INT-9313988 and by U.S. Department of Energy grant 
DE-93ER-40762.

\newpage

\centerline {\large  FIGURE CAPTIONS}

FIG. 1:~ Schematic plot of the excitation spectrum of neutron matter, 
involving $\pi^+$ and $\pi^-$ poles, one-particle--one-hole continuum, and
possible spin-isospin sound mode $\pi^+_s$. Positive (negative) isospin 
modes have the excitation 
energy $E$ equal to $+$ ($-$) $\omega$. The dashed line shows
the integration path in the sum rules, and contributions from various
excitations are denoted by blobs (poles) and the thick line (cut).
The dashed line is inclined to the vertical axis at the angle 
$\alpha = {\rm arctg} \beta$. At various values of the velocity
$\beta$ different regions of the spectral density are sampled. Note,
that $\alpha \le 45^o$. In reality, due to multi-particle--multi-hole
continua all states have finite widths, and contributions to sum
rules are collected from everywhere along the dashed lines.

\medskip

FIG. 2:~ Toy model. 
Relative contribution to the sum rules as a function of velocity
$\beta$. The isovector density is
set arbitrarily to $\rho_{I=1} = - F_\pi^2 m_\pi$.

\newpage

\thispagestyle{empty}

%

\newpage

\thispagestyle{empty}

\end{document}